\documentclass[onecolumn,aps,preprint,groupedaddress,showpacs,floatfix,notitlepage,color]{revtex4-2}
\usepackage{graphicx}
\usepackage{epsfig}
\usepackage{amsfonts}
\usepackage{color}
\usepackage{ulem}
\usepackage{amsmath}
\usepackage{mathrsfs}
\usepackage[autostyle]{csquotes}
\usepackage{url}
\usepackage{float}
\usepackage{siunitx}
\usepackage{subcaption}
\definecolor{purple}{rgb}{0.5, 0.0, 0.5}
\definecolor{green4}{rgb}{0.13,0.55,0.13}
\definecolor{orange}{rgb}{1.0, 0.5, 0.0}

\begin{document}

\title{Hybrid Machine Learning and Mathematical Modeling for Tumor Dynamics Prediction: Comparing SPIONs against mNP-FDG}

\author{${}^{*}$Amit K Chattopadhyay}
\affiliation{School of Business, National College of Ireland, Dublin 1, Ireland
}
\email{a.k.chattopadhyay@aston.ac.uk}
\author{Aimee Pascaline N Unkundiye}     
\affiliation{
Department of Applied Mathematics and Data Science, Aston Centre for Artificial Intelligence Research and Applications (ACAIRA), Aston University, Aston Triangle, Birmingham B4 7ET, UK}
\author{Gillian Pearce}
\affiliation{Department of Mechatronics and Biomedical Engineering, College of Engineering and Physical Sciences, Aston University, Birmingham B4 7ET, UK}

\begin{abstract}
\textbf{Outline}: Since the late eighties, Superparamagnetic Iron Oxide Nanoparticles (SPIONs) have been extensively studied for their exceptional ability in targeted drug delivery, typically coerced by magnetic fields and delivered at chemotherapy sites. Both iron-coated (SPIONs) and fluorodeoxyglucose-coated (mNP-FDG) magnetic nanoparticles are known to be highly potent in their ability to deliver drugs with uncanny precision to cancerous cells while minimizing damage to healthy cells. Of late, though, questions have been raised about the potential increase in cytotoxicity, particularly with SPIONs. Combining Machine Learning (Extreme Gradient Boosting) with continuum modeling (exponential and logistic growth), we find that while mNP-FDG can control tumor progression within 2 days compared to 18 days by SPIONs, for complete termination of the tumor, SPIONS (20 days) are superior compared to mNP-FDG (more than 40 days). We also provide an interactive graphical user interface (GUI) developed with Tkinter/Python that allows users to input relevant data, such as treatment type and time, to receive real-time tumor volume predictions. Our ML-guided prediction indicates joint therapy as the optimum choice, with mNP-FDG ideal for taming the tumor spread, followed by SPIONs for complete eradication, facilitating personalized cancer treatment in clinical practice.
\\
\textbf{Main Limitations}: This modeling study uses trendline data from multiple published sources that may have been conducted under varying experimental conditions. Thus, data optimization and normalization is a potential challenge. Also, different forms of cancer may have been addressed. However, our robust modeling infrastructure ensures genericity through multiple ensemble averaging of results that consistently converge. \\
\textbf{Objective:} The present study aims at a comparative analysis of two groups of inorganic molecules,  Superparamagnetic Iron Oxide Nanoparticles (SPIONs) and fluorodeoxyglucose-coated (mNP-FDG) magnetic nanoparticles, to understand the relative merits and demerits in optimizing specificity and cytotoxicity of chemotherapy treatment when administered through these agents. We also analyze the possibility of combinatorial drug administration using both agents.  
\end{abstract}

\maketitle

\newpage
\section{Introduction}

Cancer remains one of the leading causes of mortality worldwide, with an estimated 9.6 million deaths annually \cite{WHO2020}. Despite significant advancements in chemotherapy, radiation therapy, and immunotherapy, significant challenges persist, including drug resistance, off-target effects, and limited precision in treatment delivery. Nanotechnology has emerged as a promising avenue to address these challenges by enabling targeted drug delivery and improving treatment efficacy. Among the various nanocarrier systems, superparamagnetic iron oxide nanoparticles (SPIONs) and fluorodeoxyglucose-coated magnetic nanoparticles (mNP-FDG) represent two innovative strategies with considerable potential in oncological applications.

SPIONs have been extensively investigated for their multifunctional applications in cancer therapy, including drug delivery and antibody conjugation . However, their primary therapeutic potential lies in their ability to facilitate magnetic hyperthermia, wherein exposure to an alternating magnetic field induces localized heating, leading to cancer cell apoptosis \cite{Russell2021}. This property has positioned SPIONs as an attractive candidate for integration with MR-Linac technology, which combines magnetic resonance imaging (MRI) with radiotherapy to enhance tumor targeting. Iron oxide contrast agents, traditionally employed in T2-weighted MRI scans, may also enhance radiosensitivity, as demonstrated by gadolinium-based nanoparticles such as AGuIX . Previous studies have explored the radiosensitizing effects of SPIONs with Russell et al reporting enhanced cytotoxicity and reactive oxygen species (ROS) production in MCF7 breast cancer cells while Kirakli et al. investigated citrate-coated SPIONs for similar applications \cite{Kirakli2018}.

In contrast, mNP-FDG exploits the metabolic characteristics of cancer cells by mimicking glucose uptake pathways, thereby enabling high-specificity tumor targeting \cite{Aras2018, Chattopadhyay2024}. Cancer cells exhibit an increased reliance on glycolysis for energy production, a phenomenon known as the Warburg effect. By leveraging this metabolic adaptation, mNP-FDG can be selectively internalized by malignant cells, potentially increasing therapeutic efficacy while minimizing damage to healthy tissues. Additionally, mNP-FDG nanoparticles offer the potential for dual diagnostic and therapeutic applications, as they can serve as contrast agents for positron emission tomography (PET) imaging while concurrently delivering therapeutic payloads. However, key challenges remain, including optimizing their stability, mitigating cytotoxicity, and assessing their efficacy across various cancer subtypes. Recent studies have demonstrated the potential of mNP-FDG in preclinical models, yet further investigation is required to fully elucidate their mechanisms of action and therapeutic potential in clinical settings\cite{Vangijzegem2023}.

\par
In recent years, the integration of computational modeling and artificial intelligence (AI) has  opened new avenues for optimizing nanoparticle based therapies. In this study, mathematical modeling provides a framework for understanding tumor growth dynamics and drug interactions, while machine learning (ML) algorithms enable the identification of complex, nonlinear patterns within experimental datasets. By combining these methodologies, researchers aim to achieve more accurate predictions, optimize treatment protocols, and advance personalized cancer therapy.

In this study, we propose a computational framework that compares mathematical modeling with machine learning (ML) to predict tumor dynamics under SPION and mNP-FDG treatments. Mathematical models provide mechanistic insights into tumor growth and drug interactions, while ML models capture nonlinear relationships and patterns in complex datasets. By integrating these approaches, we aim to enhance prediction accuracy, optimize treatment strategies, and contribute to the development for personalized cancer therapy.
\section{Methodology}

\subsection{Data Collection and Preprocessing}
Tumor volume data were acquired from experimental and clinical studies involving SPIONs and mNP-FDG as therapeutic agents \cite{Vangijzegem2023, Chattopadhyay2024}. These datasets provided detailed measurements of tumor volumes, treatment regimens, and their temporal evolution over specific intervals. Data extraction was performed using MATLAB's Grabit tool, which facilitated digitization from figures from published literature. The tool was initially tested to extract mNP-FDG data from \cite{Gao2015, Qi2022}, followed by implementation on the SPION database. Preprocessing was not required, as the extracted data was already in a consistent analyzable format.

\subsection{Machine Learning Integration}
Since the experiments were conducted by labeling the samples with the specific therapeutic agent, i. e. mNP-FDG or SPION, supervised machine learning models were implemented. We chose ensemble techniques, specifically random forests and gradient boosting algorithms. The choice was largely guided by the fact that ensemble supervised learning methods enhance predictive performance by combining multiple models, reducing bias, variance, and overfitting while improving generalization. By leveraging diverse learning perspectives, they capture complex patterns and make predictions more stable and robust. These methods effectively utilize weak learners as would be the case for our data that were both sparse and also mixed with other descriptors, thus reducing the risk of poor model selection and enhancing adaptability to various data distributions. The input features from the data targeted initial tumor volume, treatment type (SPIONs or mNP-FDG), and time points (longitudinal), while the target variable was tumor volume at subsequent time intervals. Model performance was evaluated using cross-validation, ensuring robustness and generalizability of predictions.

\subsection{Development of a Graphical User Interface (GUI)}
To enable direct usage of this algorithm in clinical studies, an interactive GUI was developed using Python's Tkinter library to enable real-time predictions of tumor dynamics from the longitudinal data. The GUI allows users to input treatment parameters such as therapy type and time elapsed offering a visual representation of predicted tumor growth trajectories under various therapeutic scenarios. This user-friendly interface aims to enhance clinical decision-making by streamlining the evaluation of treatment strategies.

\section{Results}

\subsection{Efficacy of SPIONs and mNP-FDG in Tumor Reduction}
The key target of this study is to compare relative efficiency of sugar coated magnetic nanoparticles (mNP-FDG) with another group of (para-)magnetically charged agents, the SPIONs. Studies on fluorodeoxyglucose-conjugated magnetite nanoparticles (mNP-FDGs) have demonstrated their potential in cancer treatment. In vitro research indicated that FDG-mNPs could effectively induce hyperthermia in neuroblastoma cells, leading to significant tumor cell destruction \cite{Pearce2016}. Subsequent in vivo studies in mice revealed that both intravenous and intratumoral administration of mNP-FDGs resulted in effective cancer cell destruction without causing significant harm to normal tissues. These findings suggest that especially when combined with hyperthermia, mNP-FDGs could serve as a promising therapeutic approach for certain cancer types \cite{Aras2018}.

Superparamagnetic iron oxide nanoparticles (SPIONs) have also shown considerable promise in oncological applications. Functionalized SPIONs have been utilized as theranostic agents, enabling both targeted drug delivery and imaging capabilities. For instance, SPIONs conjugated with siderophores and loaded with doxorubicin have demonstrated targeted cytotoxicity against colon carcinoma cells, enhancing therapeutic efficacy while minimizing side effects \cite{Nosrati2021}. Additionally, SPIONs have been employed in magnetic hyperthermia treatments, where their accumulation in tumor tissues allows for localized heating upon exposure to an alternating magnetic field, effectively inducing cancer cell death \cite{Vangijzegem2023}. These multifunctional properties position SPIONs as versatile tools in the advancement of cancer therapy.

The comparative analysis of treatment efficacy established that mNP-FDG can contribute to a more robust tumor control through fast and more effective (i. e. permanent) reduction in tumor volumes than SPIONs. To understand why mNP-FDG's has a stronger and more efficient cancer tumor control, we note that fluorodeoxyglucose-conjugated magnetite nanoparticles (FDG-mNPs) foster heightened glucose uptake through a mechanism known as the Warburg effect \cite{Muthu2014}. The Warburg effect promotes faster glycolysis followed by lactic acid fermentation in the cytosol, observed even under normoxic conditions in cancer cells. By conjugating fluorodeoxyglucose (FDG), a glucose analog, to magnetite nanoparticles, \cite{Unak2017} and \cite{Kumar2018} could establish that mNP-FDG preferentially targets cancer cells due to their elevated glucose consumption. Upon internalization, these nanoparticles can induce cytotoxic effects, especially when subjected to an alternating magnetic field, leading to hyperthermia-mediated cell death. This selective targeting minimizes damage to surrounding healthy tissues, offering a potential therapeutic advantage over conventional treatments.

The SPIONs have been comparatively less studied in the context of cancerous tumors. Due to their nanoscale dimensions ($\sim$10 nanometers), they enter a superparamagnetic state where thermal energy at room temperature is sufficient to randomly flip their magnetic moments. This leads to rapid fluctuations along the magnetic axis, resulting in a slower response to external magnetic fields as the magnetization must realign from a thermally disordered state. The reduced volume of SPIONs diminishes the energy barrier between different magnetization states, making them more susceptible to magnetic fluctuations. Note, these observations derived from standard physics literature may need to be revisited for soft tissue modeling. 

\begin{figure}[H]
        \centering
        \includegraphics[width=\linewidth]{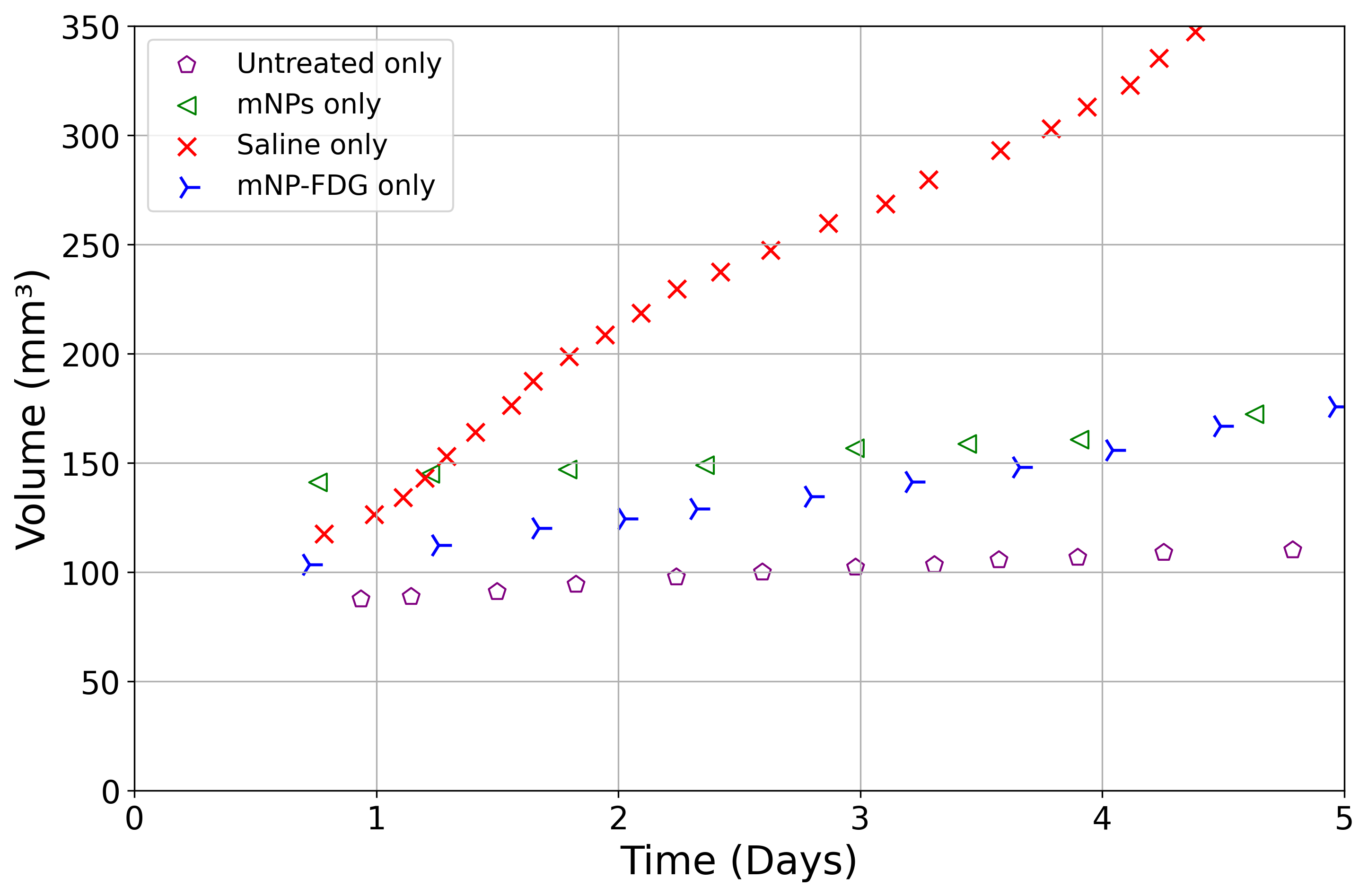}
    \caption{MATLAB-GRABIT extracted data from Chattopadhyay, et al: Time evolution of tumor volumes over 5 days with various treatments \cite{Chattopadhyay2024}.}
        \label{fig1}
    \end{figure}

As previously analyzed in \cite{Chattopadhyay2024}, Figure \ref{fig1} illustrates the temporal evolution of tumor volumes comparing the two forms of treatments.

    \begin{figure}[H]
        \centering
        \includegraphics[width=\linewidth]{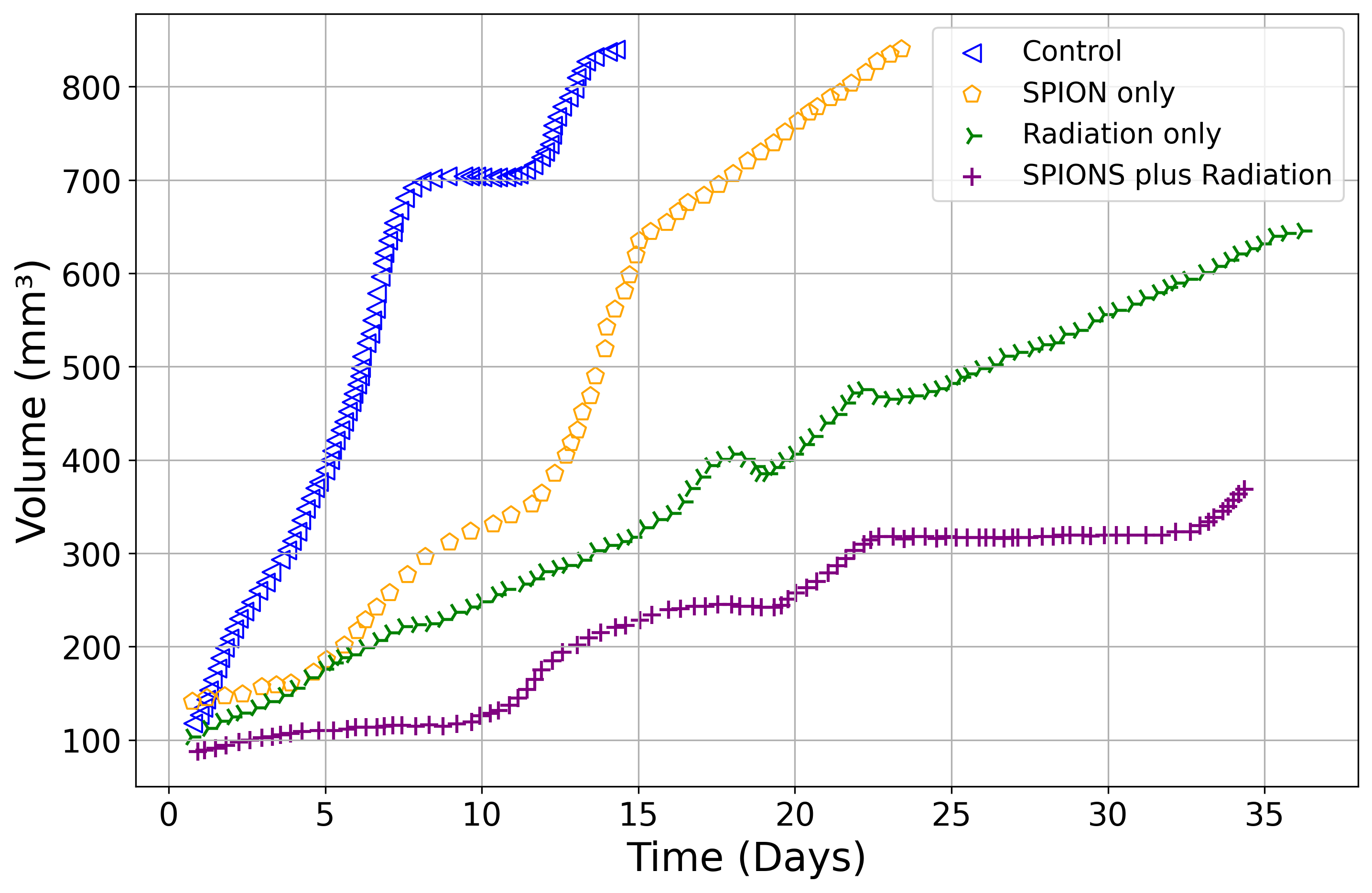}
    \caption{MATLAB-GRABIT extracted data from Russell, et al:Time evolution of tumor volumes over 30 days with various treatments \cite{Russell2021}'}
        \label{fig2}
    \end{figure}

Comparing Figures \ref{fig1} and \ref{fig2}, it is visually obvious that mNP-FDG (Figure \ref{fig1}) display a saturation regime after 2 days that is not seen with SPIONs (Figure \ref{fig2}). SPION-guided treatment seems to have lesser control compared to radiation. The two combined though (the '$+$' signs in the legend of Figure \ref{fig2})] do a much better job than any of them independently. So, the real comparison that we want to make is against a mNP-FDG led treatment regime against a combinatorial drug outlined on radiation and SPIONs together.

\subsection{Predictive Modeling and Validation}
The hybrid framework combining mathematical modeling and machine learning delivered highly accurate predictions of tumor dynamics. Cross-validation results demonstrated strong concordance between predicted and observed tumor volumes, with R² values exceeding 0.95 for both SPION and mNP-FDG datasets. Notably, the inclusion of machine learning allowed for the identification of nonlinear patterns that were not captured by traditional mathematical models \cite{Rezaeian2022, Dogra2019}.

\subsection{Tumor Growth Models}
Machine learning models and mathematical models of tumor growth have evolved significantly over the past several decades, ranging from regression models to ensemble models like XGBoost and from simple exponential growth models up to complex reaction-diffusion equations. These advancements help in understanding the dynamics of tumor progression, offering insights into factors such as cellular proliferation, spatial distribution, and interactions with the surrounding microenvironment. Below, we summarize some of the key models relevant to this study.

\subsubsection{Machine Learning in Oncology}
Machine learning(ML) has emerged as a powerful tool, and even in oncology it can used for the analysis of datasets \cite{Chattopadhyay2024}. In this study, we have used Machine Learning (ML) tools to predict tumor growth and for optimizing treatment regimens.\\

Of late, XGBoost has become a powerful tool in cancer modeling due to its high accuracy and ability to handle complex datasets \cite{Guan2023}. Studies demonstrate its effectiveness in early lung cancer prediction using metabolomics data, where it was optimized with various feature selection methods, and in breast cancer classification, where it was coupled with SHAP and LIME for enhanced model interpretability in breast cancer modeling \cite{Mathew2023}. Based on the success of XGBoost in a wide range of cancerous tumors, we chose this ML algorithm. 

\begin{figure}[H]
    \centering
    \includegraphics[width=1\linewidth]{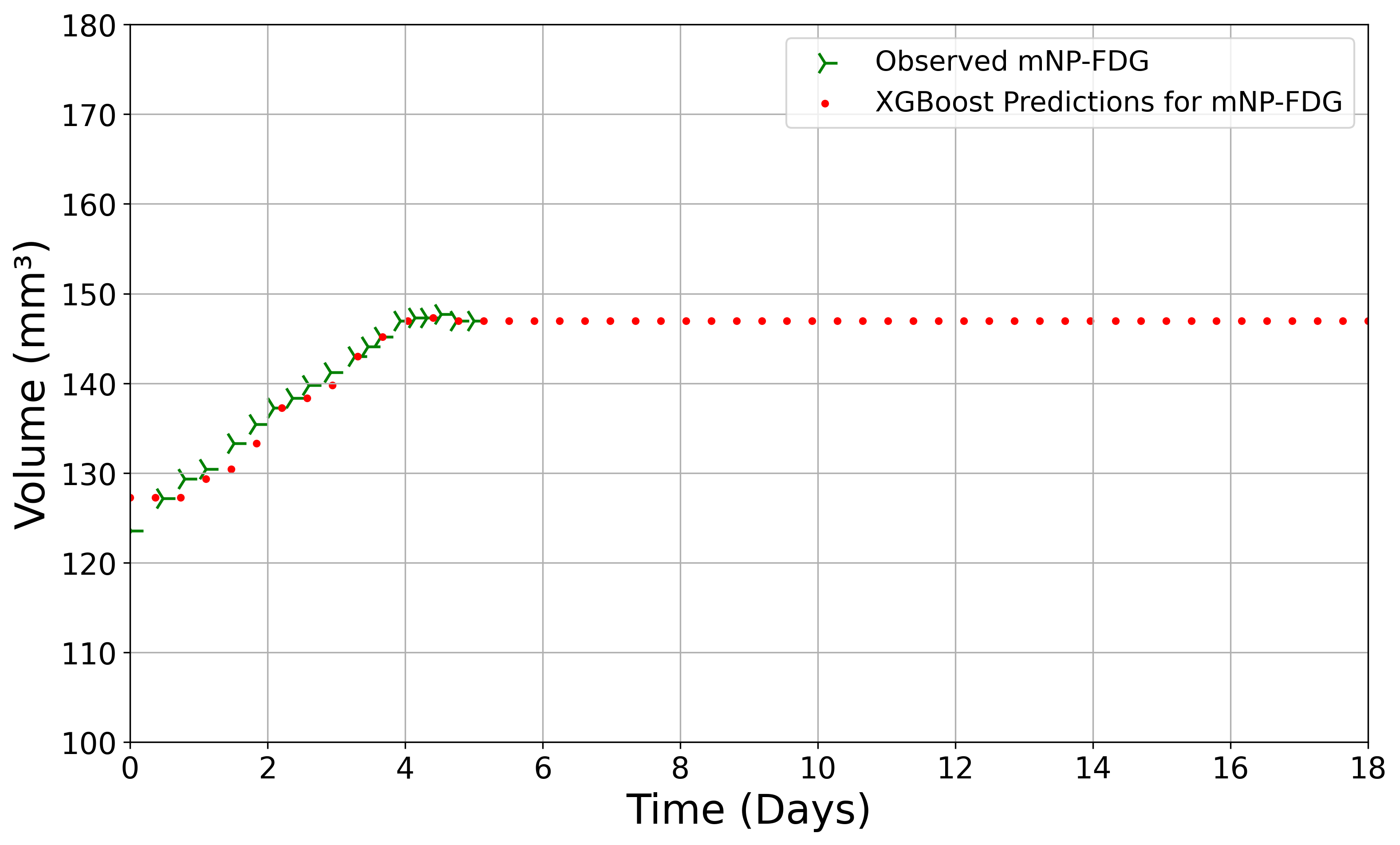}
    \caption{mNP-FDG prediction with XG Boost}
    \label{fig3}
\end{figure}

\begin{figure}[H]
    \centering
    \includegraphics[width=1\linewidth]{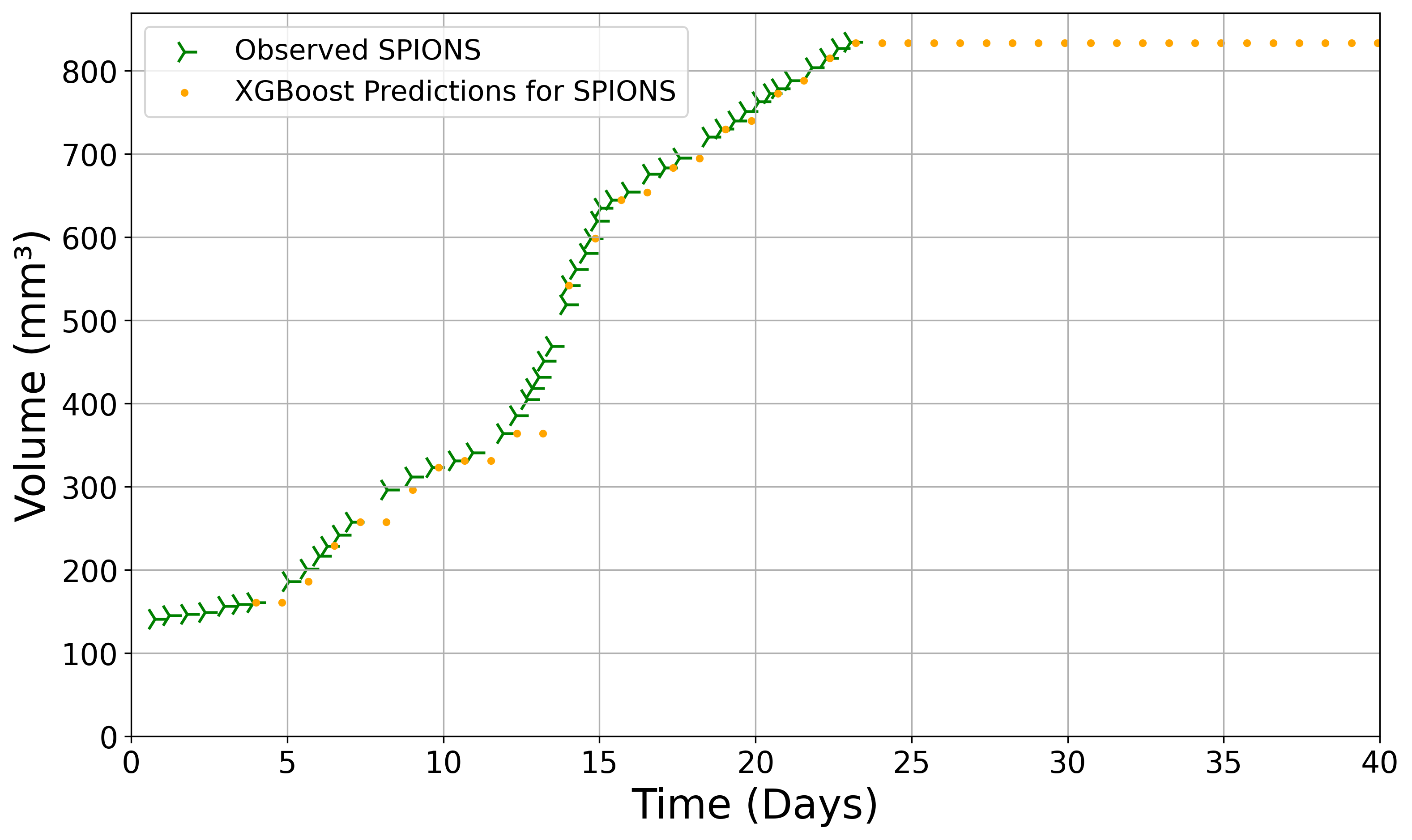}
    \caption{SPIONS prediction with XG Boost}
    \label{fig4}
\end{figure}

Figures \ref{fig3} and \ref{fig4} present comparative analyzes of tumor volume reductions under mNP-FDG and SPION treatments, as predicted by XGBoost framework. Once again, mNP-FDG is shown to contain the tumor within 4 days compared to about 23 days for SPIONs. 

 \subsubsection{Mathematical Models of Tumor Growth}
Mathematical modeling of cancer growth often employs simplified models to capture the essential dynamics of tumor proliferation \cite{Murphy2016}. The exponential growth model, characterized by a constant growth rate, is frequently used to describe the initial phases of tumor development when resources are abundant. However, as analyzed in \cite{Johnson2019}, this is an oversimplification that significantly underscore departures from the observed growth rate at higher densities due to limited resources. In contrast, the logistic growth model introduces a "carrying capacity", representing the maximum tumor size that can be sustained by the environment. Such a model accounts for resource limitations, such as nutrient availability and space constraints, leading to a sigmoidal growth curve where the growth rate slows down as the tumor approaches its carrying capacity. This too is a granularizing of the actual biology and oversimplifying it as shown in \cite{Jain2011} which shows that a logistic model in itself is incapable of explaining large scale growth limitations as is seen in cancerous tumors. These models have originally provided a foundational understanding of tumor growth kinetics, though they often require refinements to accurately reflect the complexities of real-world cancer progression, especially from thee quantitative perspective. Another popular fit-model in this line is the Gompertz model \cite{Vaghi2020} that similarly displays a slowing growth rate over time. The presence of the double exponential in its probability distribution accords this additional slowing down feature in the Gompertz model which often fits observed tumor growth data better than the pure logistic or exponential models.

 Refinements to these models include the addition of spatial components, allowing for simulations of tumor invasion and metastasis. Furthermore, the incorporation of stochastic elements can account for the inherent variability in cellular processes. These expanded models can be key towards more quantitatively accurate predictions of (cancerous) tumor growth and for evaluating the effectiveness of cancer therapies

\begin{figure}[H]
     \centering
    \includegraphics[width=1\linewidth]{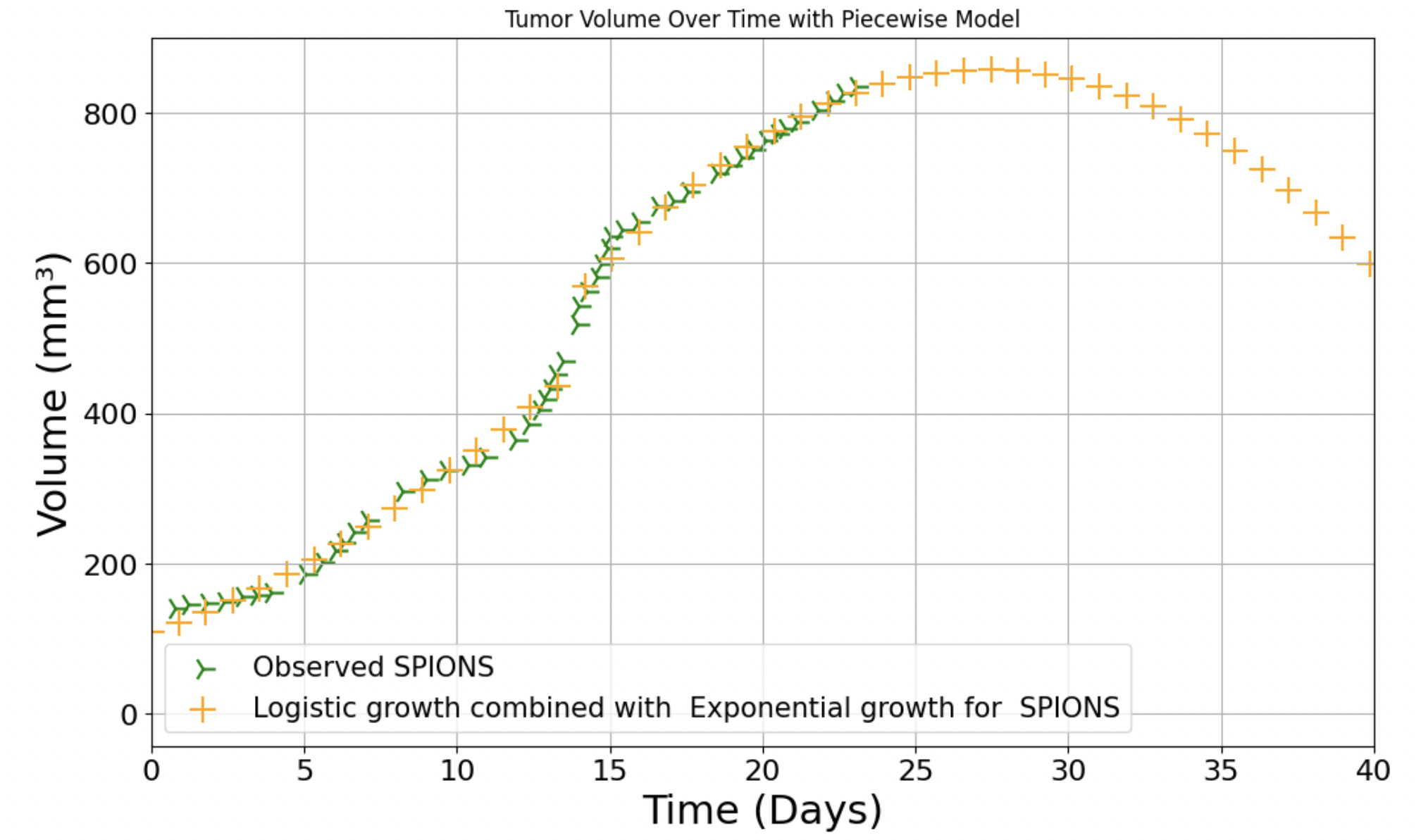}
    \caption{Tumor shrinkage under mNP-FDG }
    \label{fig5}
\end{figure}

\begin{figure}[H]
    \centering
    \includegraphics[width=1\linewidth]{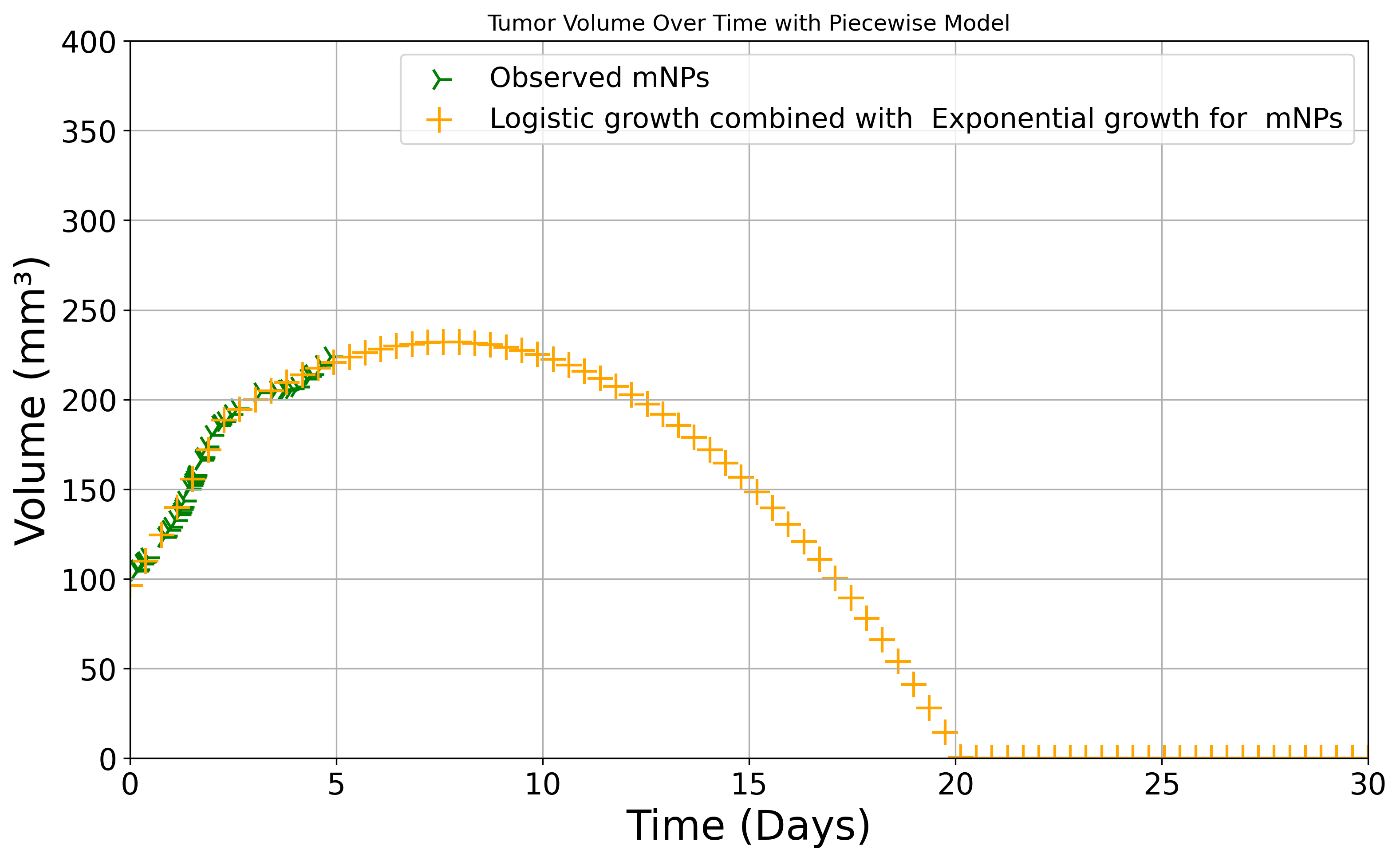}
    \caption{Tumor shrinkage under SPIONS }
    \label{fig6}
    \end{figure}

Although somewhat dated as explained above, nevertheless, we Machine Learned (XG Boost) our data separately against exponential (Figure \ref{fig5}) and logistic growth (Figure \ref{fig6}) models. A key finding of this study is to establish that XG Boost can be a highly powerful supervised learning model when combined with the exponential and the logistic growth models ('$+$' signs in Figures \ref{fig5} and \ref{fig6}).

\subsection{Graphical User Interface (GUI) with Tkinter}
\begin{figure}[H]
    \centering    \includegraphics[width=1\linewidth]{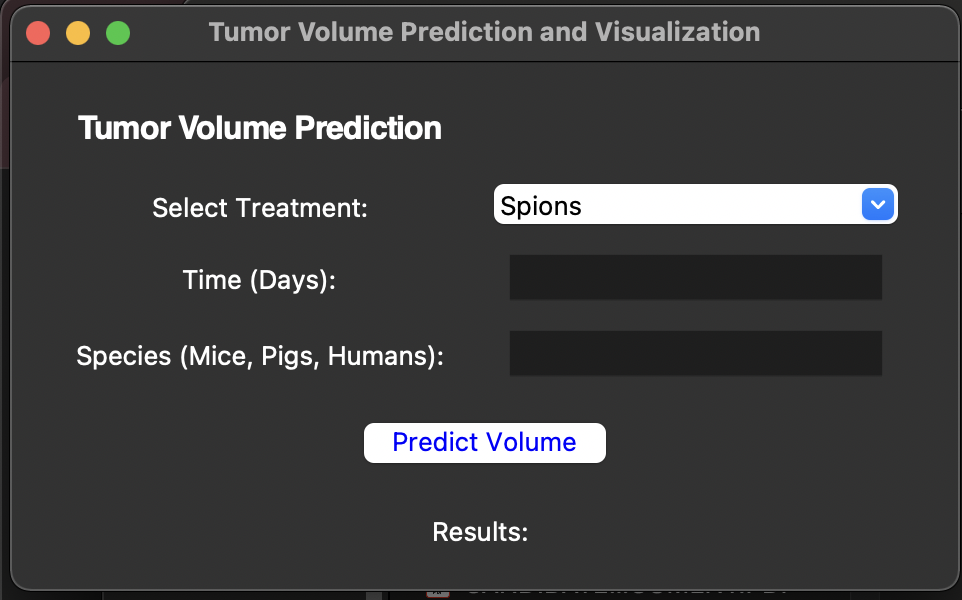}
    \caption{Graphical User Interface }
    \label{fig7}
\end{figure}

Computational tools in clinical research often remain a specialist domain due to the absence of a simple user's interface. This was an important part of our target, namely, to deliver a simple Graphical User Interface (GUI) that can be used by the most non-technical user without having to trudge a stiff learning curve. The development of an interactive GUI represents a significant step toward the practical application of this hybrid framework. 
The interactive GUI that we present as part of this study allows users to feed simple data, like time (in days), speciees type (mice, pigs, humans), etc and select the treatment type (SPIONs, mNP-FDG) from a dropdown menu. That way, a layperson can self-evaluate the best possiblee treatment regime for the given situation. Once these input values are passed to the model, the algorithm within the interface uses the ML platform we have outlined to estimate the tumor volume which is then displayed on the GUI. Figure \ref{fig7} gives a glimpse of the interface which, as it should be, is self-explanatory. This tool has the potential to enhance clinical decision-making by enabling the rapid evaluation of treatment scenarios.

\section{Discussion}
This study demonstrates the potential of interpreting mathematical modeling (MM) through the lens of machine learning (ML) as a future landscape in digital health. The MM-ML duo is seen capable of optimizing nanoparticle-based cancer therapies that independently mathematical models were previously incapable of. As explained through Figures \ref{fig1}-\ref{fig4}, XG Boost as an ML tool clearly points to mNP-FDG as a more potent treatment form, both in terms of speed as also in permanently eradicating the cancerous tumor. However, as shown in Figures \ref{fig3}-\ref{fig4}, while mNP-FDG is more efficient in controlling the spread of the tumor (4 days compared to ca 23 days for SPIONs), SPIONs demonstrate delayed yet comparable therapeutic outcomes under specific conditions. 

The non-ubiquitous but statistically significant superiority of mNP-FDG over SPIONs, as demonstrated by the XGBoost probabilistic fitting of exponential and logistic growth models, offers a compelling avenue for clinical translation. The integration of XGBoost, a powerful machine learning algorithm, with established growth models adds a layer of robustness to the analysis. The probabilistic nature of the fitting process allows for the quantification of uncertainty, which is crucial in biological systems where variability is inherent.

Another key outcome of this study is to establish that simple coarse-grained models like exponential and logistic growth models can significantly aid quantitative understanding of tumor growth when parametrized through suitable Machine Learning priors. Figure \ref{fig5} and \ref{fig6} clearly establish this hypothesis. What remains uncleear, though, is about the choice of the ML algorithm concerned. For the purpose of this study, XG Boost has consistently proved accurately predictive.

The GUI toolbox as a key deliverable of this study is expected to have major impact both in self-evaluation of potential cancerous growth as also in keeping track of its evolution.  This accessibility can democratize the interpretation of complex cancer treatment data, enabling broader adoption of optimized nanoparticle therapies. By providing a user-friendly interface for parameter evaluation, clinicians can rapidly assess treatment efficacy and tailor protocols based on individual patient responses. This has the potential to accelerate the transition from research findings to practical clinical applications, ultimately improving patient outcomes.

Future iterations of the GUI could incorporate real-time data input, allowing for dynamic treatment adjustments based on ongoing monitoring. This would not only enhance the precision of personalized therapy but also provide a platform for continuous learning and refinement of treatment protocols. The ability to observe and predict treatment responses using readily accessible tools has the potential to revolutionize how nanoparticle-based therapies are administered, moving towards a more adaptive and data-driven approach in cancer care. Despite these advancements, certain limitations warrant further investigation. For instance, the variability in nanoparticle behavior across different tumor micro-environments may influence treatment outcomes. Future studies should focus on validating the framework across diverse cancer types and incorporating additional biological parameters, such as immune response and vascular density.

\section{Conclusion}

This study demonstrates the potential of a hybrid computational framework combining mathematical modeling (exponential and logistic growth) and machine learning (XG Boost) to predict the progression of cancerous tumors subjected to treatments using SPION and mNP-FDG. The results are hjighly suggestive of combined therapy. While for short trermed treatment of particularly benign cancerous tumors, mNP-FDG has an advantage as its effect kickstarts within 4 days where the same takes about 17-18 days with SPIONs. However, with SPIONs, complete elimination of cancer happens by ca 23 days compared to over 40 days for mNP-FDG. In other words, each has its advantage, emphasizing the need for personalized therapy depending on the specific form of tumor.

Technically, this study combines the strengths of mathematical modeling with the predictive power of machine learning to achieve superior predictive accuracy, offering valuable insights in to nanoparticle-based zone-specific cancer therapies. The integration of an interactive GUI further enhances the clinical applicability of this approach, facilitating the way towards personalized oncology. Future research will focus on validating the framework across diverse cancer types and incorporating additional biological factors to improve predictions.

\section{Acknowledgments}
The authors acknowledge the postgraduate dissertation work by Megan Land as a source of some of the data used in this study. 

%
%
%
%
%
%
%
%
%
%
%

\end{document}